\documentclass{iopart}

\usepackage{latexsym}
\usepackage{amssymb}
\usepackage{epsfig}

\begin{document}

\newcommand{\muB}{\hat{\mu}}



\newcommand{\connE}{\ensuremath{\tilde{\nabla}}}
\newcommand{\phib}{\ensuremath{\bar{\phi}}}
\newcommand{\connM}{\ensuremath{\nabla}}

\newcommand{\conngam}{\ensuremath{\;^{(\tilde{\gamma})}\tilde{\nabla}}}
\newcommand{\conneta}{\ensuremath{\bar{\nabla}}}
\newcommand{\conng}{\ensuremath{\tilde{\nabla}}}

\newcommand{\conngamt}{\ensuremath{\;^{(\gamma)}\nabla}}
\newcommand{\connetat}{\ensuremath{\hat{\nabla}}}
\newcommand{\conngt}{\ensuremath{\nabla}}

\newcommand{\nablaM}{\ensuremath{\nabla}}
\newcommand{\nablaE}{\ensuremath{\tilde{\nabla}}}
\newcommand{\nablaS}{\ensuremath{\hat{\nabla}}}

\newcommand{\metE}{\ensuremath{\tilde{g}}}
\newcommand{\metM}{\ensuremath{g}}
\newcommand{\RWmetE}{\ensuremath{\tilde{\gamma}}}
\newcommand{\RWmetM}{\ensuremath{\gamma}}
\newcommand{\etaE}{\ensuremath{\tilde{\eta}}}
\newcommand{\etaM}{\ensuremath{\eta}}
\newcommand{\permetE}{\ensuremath{\tilde{h}}}
\newcommand{\permetM}{\ensuremath{h}}

\newcommand{\volE}{\ensuremath{\sqrt{-\metE}}}
\newcommand{\volM}{\ensuremath{\sqrt{-\metM}}}
\newcommand{\RiemE}{\ensuremath{\tilde{R}}}
\newcommand{\RiemM}{\ensuremath{R}}
\newcommand{\EinE}{\ensuremath{\tilde{G}}}
\newcommand{\EinM}{\ensuremath{G}}

\newcommand{\Vp}{\ensuremath{\frac{dV}{d\mu}}}

\newcommand{\bim}{\ensuremath{B}}
\newcommand{\ibim}{\ensuremath{C}}

\newcommand{\grad}{\ensuremath{\vec{\nabla}}}

\newcommand{\metS}{\ensuremath{\hat{g}}}
\newcommand{\AS}{\ensuremath{\hat{A}}}
\newcommand{\GS}{\ensuremath{\hat{G}}}
\newcommand{\KS}{\ensuremath{\hat{K}}}
\newcommand{\SeS}{\ensuremath{\hat{S}}}
\newcommand{\JS}{\ensuremath{\hat{J}}}
\newcommand{\cS}{\ensuremath{\hat{c}}}

\newcommand{\KE}{\ensuremath{\tilde{K}}}
\newcommand{\aE}{\ensuremath{\tilde{A}}}
\newcommand{\cE}{\ensuremath{\tilde{c}}}

\pacs{98.5}
\submitto{\CQG}

\paper{The Geometry Of Modified Newtonian Dynamics}
\author{Constantinos Skordis}
\address{School of Physics and Astronomy, University of Nottingham, Nottingham, NG7 2RD, United Kingdom.}
\ead{skordis@nottingham.ac.uk}
\author{Tom Zlosnik}
\address{Perimeter Institute for Theoretical Physics, 31 Caroline street North, Waterloo, Ontario N2L 2Y5, Canada}
\ead{tzlosnik@perimeterinstitute.ca}

\begin{abstract}
Modified Newtonian Dynamics is an empirical modification to Poisson's equation which has had success in accounting
for the `gravitational field' $\Phi$ in a variety of astrophysical systems. The field $\Phi$ may be interpreted in terms of the weak
field limit of a variety of spacetime geometries. Here we consider three of these geometries in a more comprehensive
manner and look at the effect on timelike and null geodesics. In particular we consider the Aquadratic Lagrangian  (AQUAL) theory, Tensor-Vector-Scalar (TeVeS)  theory
and Generalized Einstein-{\AE}ther (GEA) theory.

  We uncover a number of novel features, some of which are specific to the theory considered while others are generic. 
In the case of AQUAL and TeVeS theories, the spacetime exhibits an excess (AQUAL) or deficit (TeVeS) solid angle akin to the case of a Barriola-Vilenkin global monopole.
In the case of GEA, a disformal symmetry of the action emerges in the limit of $\grad\Phi\rightarrow 0$.
Finally,
in all theories studied, massive particles can never reach spatial infinity while photons can do so only after experiencing infinite redshift.
\end{abstract}

\section{Introduction}

Einstein's theory of General Relativity is a supremely successful theory on scales of the solar system and below. Although specific predictions about such diverse phenomena as the gravitational redshift of light, energy loss from binary pulsars, the rate of precession of the perihelia of closed orbits, and light deflection by the sun are not unique to General Relativity, it must be regarded as highly significant that General Relativity is  consistent with each of these tests and more. 

The success of General Relativity on larger scales still is perhaps less clear. On the largest observable scales, observations appear to have converged upon a simple model: the so-called Concordance Model. The addition of a cosmological constant and an approximately pressureless fluid (termed Cold Dark Matter) to a system of General Relativity coupled to the fields of the standard model of particle physics (and in conjunction with appropriate initial conditions) appears sufficient to give an excellent account of observables such as the large scale structure in the universe and the cosmic microwave background anisotropies. 

Despite the success of the Concordance Model, the requirement of the existence of an extra, stable matter field as yet unobserved in any small scale experiment and the accompanying requirement of a (by all conventional standards) very small but non-zero cosmological constant has led some to question whether there is something lacking about the model. A differing model of the universe at the level of linear cosmological perturbations would presumably amount to a different account at the level of nonlinear perturbations: i.e. on small scales such as galaxies or the solar system.
Crucially though, galactic scales do indeed exhibit mass discrepancies: stars move as if pulled on than a mass greater than the observed luminous mass, gravitational lensing by many galaxies is greater than one would expect assuming General Relativity and the observed luminous mass.
Currently it is not entirely clear that they are of the precise nature predicted by the Concordance Model. In light of the uncertainty about what constitutes 
the `dark sector' in the universe, and the possible limitations of the Concordance Model, one may instead attempt a phenomenological description of the mass
 discrepancies on small scales and attempt to work from there to an appropriate cosmological model.

Indeed, in 1983 it was realized by Milgrom \cite{Milgrom1983a} that the emerging evidence for the presence of dark matter in galaxies
could rather follow from a modification either to how `baryonic' matter \footnote{Baryonic matter is defined to be matter which consists of ingredients from the standard model of particle physics} responded to the Newtonian gravitational field it created
or to how the gravitational field was related to the baryonic matter density. Collectively these ideas are referred to as MOdified Newtonian Dynamics (henceforth MOND).  By way of illustration, MOND may be considered as a modification to the nonrelativistic Poisson equation~\cite{BekensteinMilgrom1984}:
\begin{equation}
\label{eq:mond}
\grad\cdot\left[  \mu\left(\frac{|\grad\Phi|}{a_{0}}\right)\grad\Phi\right]= 4\pi G_N\rho
\end{equation}
where $\Phi$ is the gravitational potential, $\grad$ is the grad operator, $a_{0}$ is a number with dimensions Length$^{-1}$, $G_N$ is Newton's constant and $\rho$ is the baryonic matter density. The number $a_{0}$ is determined by looking at the dynamics
of visible matter in galaxies and a value of the order of the inverse Hubble length seems to be preferred \cite{SandersMcGaugh2002}. The function $\mu(x)$ would simply be equal to unity in Newtonian gravity. In MOND, the functional form is only fixed at its limits: $\mu \rightarrow 1$ as $x \rightarrow \infty$ and $\mu \rightarrow x$ as $x \rightarrow 0$. We will call the limit $\mu\rightarrow x$ the `modified gravity' (MG) regime. 

The success of (\ref{eq:mond}) may simply reflect the behaviour of dark matter on these scales.
Alternatively, the complete absence of anything other than a gravitational potential and baryonic matter from this equation may appear to indicate that the origin of (\ref{eq:mond}) lies along the lines of a modification to how the baryonic matter affects spacetime i.e. a modification to gravity.

Considerable work has gone into exploring theoretical implications
 (\cite{LueStarkman2004,BekensteinMagueijo2006,BradaMilgrom1994b,MalekjaniRahvarHaghi2008,DaiMatsuoStarkman2008a,MilgromSanders2007,MatsuoStarkman2009}) 
and observational implications (\cite{deBlokMcGaugh1998,FamaeyBinney2005,SandersMcGaugh2002,Zhao2005,AngusFamaeyZhao2006,SamurovicCirkovic2008,LondrilloNipoti2008,Gentile2008}) of equation (\ref{eq:mond}).
If MOND is taken to be a modification to gravity, it would seem necessary to find a theory where (\ref{eq:mond}) emerges from a relativistic framework, or a wider framework wherein at least the successful predictions of General Relativity could be recovered in the appropriate limit. In this absence of this, it is unclear precisely when one is able to use MOND and when it is appropriate to compare the predictions of MOND with those of the dark matter paradigm. 
This was quickly recognized and in 1984 Bekenstein and Milgrom introduced a generally covariant theory which could reproduce MOND phenomenology in the absence of dark matter \cite{BekensteinMilgrom1984}. We refer to this theory as AQUAL (not to be confused with a nonrelativistic Lagrangian theory sometimes given the same name), named after the AQuadratic Langragian of the theory. 

Since then, a variety of other relativistic models have been proposed which seek to reproduce aspects of equation  (\ref{eq:mond}) in the quasistatic, weak field limit, (see for instance 
\cite{SoussaWoodard2003,Sanders2005,Bekenstein2004a,NavarroVanAcoleyen2006,ZlosnikFerreiraStarkman2007,BlanchetTiec2008b,BrunetonEtAl2008,Milgrom2009,Skordis2009a,Skordis2009b,ZhaoLi2010}). 
Of these, we single out two theories which we will further use in this work, namely Bekenstein's Tensor-Vector-Scalar (TeVeS) theory~\cite{Bekenstein2004a} and the generalized 
Einstein-{\AE}ther(GEA) theory~\cite{ZlosnikFerreiraStarkman2007}. 
A feature common to all these models is that the transition from Newtonian gravity to modified gravity is accomplished via the introduction 
of a rather unwieldy function into the model's action which is perhaps an indication that a different approach is called for.
Furthermore, none of the models can claim to be as successful as the Concordance Model on cosmological scales \cite{SkordisEtAl2006,ZuntzEtAl2010,FerreiraSkordisZunkel2008}.

In this paper we adopt a more conservative viewpoint: It may be that if there is underlying `new physics' to the MOND paradigm, it may not be convenient
to express its equations in the form (\ref{eq:mond}). We will simply assume that the new physics allows for a modified gravity regime which for appropriate sources is described at the largest distances from the source by a static, spherically symmetric spacetime metric. We are interested in the properties of this asymptotic spacetime. For instance, this spacetime may be expected to approximate that due to a black hole or a massive bound object at sufficiently large distances. This approach will enable us to deduce general properties of these  spacetimes and need not rely on the precise details of how the transition from General Relativity to modified gravity is accomplished or the details of the underlying theory.

\section{The three relativistic MOND theories}
We consider three distinct relativistic MOND theories. For each theory, we will provide the action and a brief description regarding the form of the free function involved, followed by the relation of its parameters to the
non-relativistic limit. We will then provide the spherically symmetric, vacuum field equations and proceed to solve them in the limit of large distance from the source. 
This is the limit where the 'modified gravity' regime is appropriate.

In all cases we adopt a static and spherically symmetric metric. In most of the cases we employ isotropic coordinates $\{t,r,\theta,\varphi\}$ for which the metric takes the form
\begin{equation}
ds^2 = -e^{2\Phi} dt^2 + e^{2\Psi} dL^2 
\end{equation}
where $dL^2 = \gamma_{ij} dx^i dx^j= dr^2 + r^2 \left( d\theta^2 + \sin^2\theta d\varphi^2\right) $ is a flat metric in spherical coordinates and $\Phi(r)$ and $\Psi(r)$ are two functions of $r$ which are to be determined by the field equations
for each theory. We note that both $\Phi$ and $\Psi$ are determined up-to a constant, i.e. the transformation $\Phi\rightarrow \Phi + \Phi_0$ and $\Psi \rightarrow \Psi + \Psi_0$ for constants $\Phi_0$ and $\Psi_0$ can
be undone by a further coordinate rescaling of $t$ and $r$.
 As we show further below, the metric for the three theories in the modified gravity limit, takes the form
\begin{eqnarray}
\label{met_A} \mbox{AQUAL : } \quad ds^2_{A} &=& -\left(\frac{r}{r_{0}}\right)^{2c}  dt^{2}+ \left(\frac{r}{r_{0}}\right)^{2c} dL^2  \\
\label{met_T} \mbox{TeVeS : } \quad  ds^2_{T} &=& -\left(\frac{r}{r_{0}}\right)^{2c} dt^{2}+\left(\frac{r}{r_{0}}\right)^{-2c} dL^2 \\
\label{met_GEA}\mbox{GEA : } \quad  ds^2_{\AE} &=& -\left[1+\frac{c}{3}\ln\left(\frac{r}{r_{0}}\right)\right]^{6}dt^{2} +  \frac{dL^2}{\left[1+\frac{c}{3}\ln\left(\frac{r}{r_{0}}\right)\right]^{6}}
\end{eqnarray}
for a dimensionless number $c$ and a scale $r_{0}$  emerging as an integration constant which can be chosen to roughly mark the onset of the  modified gravity regime.

The use of isotropic coordinates makes it straightforward to determine the non-relativistic limit by simply expanding $e^{2\Phi} \rightarrow 1 + 2 \Phi + \ldots$ and $e^{2\Psi}\rightarrow 1 + 2\Psi + \ldots$.
For all the three cases above, we find that $\Phi =  c \ln\frac{r}{r_0}$ which is precisely the MOND potential.
This enables us to identify  the dimensionless number $c$  with $\sqrt{G_N Ma_{0}}$ ($M$ being the mass of the source). To get a feeling for the order of magnitude of this number 
we can write it as $c \approx 1.4 \times 10^{-12} \sqrt{\frac{M}{M_\odot}}$, where $M_\odot$ is the solar mass~\cite{SandersMcGaugh2002}. 
This is a tiny number by all standards, except when $M$ starts to approach
the total mass within our Hubble volume. The non-relativistic limit is thus expected to hold for most cases of interest. It is still desirable however to look at the full 
relativistic metric.
For instance, while $\Phi$ is the same for all three theories (as required for reproducing the MOND limit), the potential $\Psi$ can be different. 
In particular AQUAL gives $\Psi = \Phi$ while TeVeS and GEA give $\Psi = - \Phi$. This is a result that could not have arisen from (\ref{eq:mond}). 
As we shall see further below, novel features can arise by the use of the full relativistic solution.

\subsection{Aquadratic Lagrangian (AQUAL) gravity}
The simplest of the theories we consider is the AQUAL theory of Bekenstein and Milgrom~\cite{BekensteinMilgrom1984}. The theory depends on an Einstein metric $\tilde{g}_{ab}$ and a scalar field $\phi$. The physical metric 
$g_{ab}$ is given by a conformal transformation $g_{ab} = e^{2\phi} \tilde{g}_{ab}$. The action is 
\begin{equation}
S_A[\tilde{g}^{ab},\phi] = \frac{1}{16\pi G} \int d^4x \sqrt{-\tilde{g}} \left[ \tilde{R}  - \frac{1}{\ell_0^2} f(X) \right] + S_{\mbox{matter}}[g]
\end{equation}
where $\ell_0$ is a scale and where
\begin{equation}
 X = \ell_0^2 \tilde{g}^{ab} \nabla_a \phi \nabla_b \phi
\end{equation}
The theory depends on a free function  $f(X)$ which must have appropriate limits to recover Newtonian or MOND gravity in the non-relativistic case. 
In particular, the non-relativistic limit of the theory gives back the MOND equation (\ref{eq:mond})
with $\mu = \frac{f_X}{2 + f_X}$ where $f_X = \frac{df}{dX}$. Defining $\lim_{X\rightarrow\infty} f(X) = f_0X$ and $\lim_{X\rightarrow 0} f(X) = f_1X^{3/2}$ we find that the Newtonian limit fixes the Newton's constant 
$G_N = \frac{2+f_0}{f_0} e^{2\phi_c} G$ where $\phi_c$ is the background (cosmological) value of $\phi$ and gives the Parameterized Post-Newtonian(PPN) parameter $\gamma = \frac{f_0-2}{f_0+2}$. In the
 MOND limit, we find that Milgrom's constant $a_0$ is given by $a_0 = \frac{4f_0}{3 f_1(2+f_0)}  e^{-\phi_c} \frac{1}{\ell_0}$.

We find the field equations assuming isotropic coordinates. The Einstein metric is given by $\tilde{g}_{00} = -e^{2\tilde{\Phi}}$ with $\Phi = \tilde{\Phi}+\phi$ and 
$\tilde{g}_{ij} = e^{2\tilde{\Psi}}\gamma_{ij}$ with $\Psi = \tilde{\Psi}+\phi$. 
The vacuum scalar field equation for our metric ansatz is
\begin{equation}
\frac{d}{dr} \left[ e^{\tilde{\Phi}+\tilde{\Psi}} r^2  f_X   \frac{d\phi}{dr} \right]  = 0
\end{equation}
where $X =  \ell_0^2 e^{-2\tilde{\Psi}}{\phi'}^2$,
while the Einstein equations are
\begin{eqnarray}
  2\tilde{\Psi}''  + \tilde{\Psi}' \left(\tilde{\Psi}' + \frac{4}{r}  \right) 
 =  -\frac{1}{2\ell_0^2} f  e^{2\tilde{\Psi}}
\\
 \frac{2}{r}\left(\tilde{\Phi}'  +\tilde{\Psi}'\right)  +  \tilde{\Psi}'\left(2\tilde{\Phi}'  + \tilde{\Psi}'\right)
  = \frac{1}{\ell_0^2}   e^{2\tilde{\Psi}}\left( X f_X  - \frac{1}{2} f \right)
\\
 \tilde{\Phi}''+ \tilde{\Psi}''  +  \frac{1}{r}\left(\tilde{\Phi}'  +\tilde{\Psi}'\right)
+ (\tilde{\Phi}')^2  
  =  - \frac{1}{2\ell_0^2} f  e^{2\tilde{\Psi}}  
\end{eqnarray}
In the MG limit $f =  f_1 X^{3/2}$ so the scalar field equation integrates once to
\begin{equation}
 \frac{d\phi}{dr}   = \sqrt{\frac{2C_0}{3f_1 \ell_0}}\frac{e^{-\frac{\tilde{\Phi}}{2}}}{r} = \frac{c e^{-\frac{\tilde{\Phi}}{2}}}{r}
\label{eq_phi_prime}
\end{equation}
where $C_0$ is an integration constant, related to the mass of the source. As $r\rightarrow \infty$ the RHS of the Einstein equations goes as
$X^{3/2} e^{2\tilde{\Psi}} \rightarrow e^{-3\tilde{\Phi}/2 - \tilde{\Psi}}\frac{1}{r^3}  $ which will decay faster than any of the terms in the LHS. This
means that in the MG limit as  $r\rightarrow \infty$, the Einstein-frame metric must be a solution to the vacuum Einstein equations and for a static spherically symmetric spacetime
that solution is unique; it is the Schwarzschild solution. We may thus write $\tilde{\Phi} = \frac{\Phi_1}{r} + \ldots$ and $\tilde{\Psi} = -\frac{\Phi_1}{r} + \ldots$, while
$\phi' = \frac{c}{r}\left(1 - \frac{\Phi_1}{2r} + \ldots\right)$ where $c =  \sqrt{\frac{2C_0}{3f_1 \ell_0}}$. The last relation integrates to
$\phi = c \ln \frac{r}{r_0}   +  \frac{c\Phi_1}{2r} + \ldots$, where the integration constant is ignored as it is irrelevant. This gives the physical metric as
\begin{eqnarray}
ds^2 &=& \left(\frac{r}{r_0}\right)^{2c} \bigg[ - \left(1 + \frac{(2+c)\Phi_1}{r} + \ldots \right)   dt^2 
  \nonumber 
\\ 
&&
\qquad \qquad \ \ \ \ 
 +   \left(1 - \frac{(2-c)\Phi_1}{r} + \ldots\right)  dL^2 \bigg]  
\label{aqual_metric_sol}
\end{eqnarray}
which is the metric (\ref{met_A}) in the limit $r \rightarrow \infty$. 

A comment is in order. In constructing (\ref{eq_phi_prime}) we have retained the positive sign when taking the square root. 
The reason is that $c<0$ is unphysical. To see this, we take the non-relativistic limit of (\ref{aqual_metric_sol}) which gives $\Phi = - |c|\ln (r/r_0)$. This implies
that the acceleration $\vec{a} = -\grad\Phi = |c|\frac{\hat{r}}{r}$ points away from the centre which would render gravity being repulsive (this is the analogue of the
negative-mass Schwarzschild solution).

It is instructive to find the physical metric in  Schwarzschild  coordinates $\bar{r}$. The coordinate transformation in the $r\rightarrow \infty$ limit implies
that $\frac{\bar{r}}{r_0} =  \left(\frac{r}{r_0}\right)^{1+c}    \left(1 - \frac{c\Phi_1}{2r} + \ldots\right)$ which can be inverted to give
$\frac{r}{r_0} = \left(\frac{\bar{r}}{r_0}\right)^{\frac{1}{1+c}}\left[1 + \frac{c\Phi_1}{2(1+c)r_0} \left(\frac{r_0}{\bar{r}}\right)^{\frac{1}{1+c}} + \ldots\right]$,
therefore $\bar{r}\rightarrow \infty$ as $r\rightarrow \infty$.
Thus we find the metric in  Schwarzschild  coordinates as 
\begin{equation}
ds^2 =
 -   \left(\frac{\bar{r}}{r_0}\right)^{\frac{2c}{1+c}} dt^2
+ \frac{   d\bar{r}^2 }{(1+c)^2}
+ \bar{r}^2  d\Omega^2 
\end{equation}
The factor $\frac{1}{1+c}$ appearing in the $d\bar{r}$ part of the metric is quite important. It implies an excess solid angle given by $1+c$. In fact, the
spatial metric is of exactly the same form as the Barriola-Vilenkin global monopole solution~\cite{BarriolaVilenkin1989} (the original solution has a deficit angle) 
so this MONDian spacetime can be thought as a warped  Barriola-Vilenkin global monopole.

The divergence of the conformal factor appearing in the metric may be a source of concern. However, no curvature singularity arises.
The  curvature tensors go as
$R^{ab}_{\phantom{ab}cd} =  P^{ab}_{\phantom{ab}cd} r^{-2(1+c)}$,
$R^{a}_{\phantom{a}b} =  P^{a}_{\phantom{a}b}  r^{-2(1+c)} $
$R = -6c(1+c) r^{-2(1+c)}$, where the $P$-tensors are constant.
So forming invariants by contracting various types of curvature tensors together, or their derivatives, we would get that a typical invariant would be
$I_{n,d} = \frac{J}{r^{2n(1+c) +d}} $, where $J$ is a real number, $n$ is the number of curvature tensors involved
and $d$ is the number of derivatives acting on the curvature tensors. Thus all curvature invariants vanish as
$r\rightarrow \infty$.

\subsection{Tensor-Vector-Scalar (TeVeS) gravity}
Bekenstein's TeVeS theory combines the AQUAL theory with a unit-timelike vector field $A_a$ considered by Sanders~\cite{Sanders1997} in the MOND framework on phenomenological grounds. The relation between the Einstein and physical metric
is changed to $g_{ab} = e^{-2\phi} \tilde{g}_{ab} - \sinh(2\phi) A_a A_b$, the so-called disformal transformation. We shall adopt the diagonal-frame formulation of the theory (see appendix of~\cite{Skordis2009b})
where the scalar field is decoupled from the vector field. The action takes the form
\begin{eqnarray}
S_T[\tilde{g},\phi,A]  &=&  S_A[\tilde{g},\phi] - \frac{1}{16\pi G} \int d^4x \sqrt{-\tilde{g}} \bigg[ K^{abcd} \tilde{\nabla}_a A_b\; \tilde{\nabla}_c A_d
\nonumber 
\\
&&
\ \ \ \
-  \lambda (A^a A_a +1) \bigg]
\label{teves_action}
\end{eqnarray}
where  $A^a = \tilde{g}^{ab}A_b$ and the tensor $K^{abcd}$ is  given by
\begin{eqnarray}
K^{abcd} &=& c_1 \tilde{g}^{ac} \tilde{g}^{bd} + c_2 \tilde{g}^{ab} \tilde{g}^{cd} + c_3 \tilde{g}^{ad} \tilde{g}^{bc}  + c_4 A^a A^c \tilde{g}^{bd},
\label{K_ten}
\end{eqnarray}
for a set of constants $\{c_1\ldots c_4\}$. This formulation of the theory is more general
 (corresponding to the proposal in~\cite{Skordis2008a}) than the one given by Bekenstein which corresponds to $c_1 = 2K - \frac{1}{4} $, 
$c_2 = -\frac{1}{2}$, $c_3= -2K + \frac{3}{4}$ and $c_4 = K - \frac{1}{4}$, where $K$ is Bekenstein's constant in~\cite{Bekenstein2004a}.
As it happens,  spherically symmetric solutions depend only on the combination $c_1-c_4$ and not on $c_2$ or $c_3$. 
Without loss of generality we shall therefore set $c_1 - c_4 = K$ from now on.

Once again, the theory depends on a free function  $f(X)$ which must have appropriate limits to recover Newtonian or MOND gravity in the non-relativistic case.
In particular, the non-relativistic limit of the theory gives back the MOND equation (\ref{eq:mond})
with $\mu =  \left(1+ \frac{2 - K }{f_0} \right)  \frac{f_X}{2 - K + f_X}$. Defining $ \lim_{X\rightarrow\infty} f(X) = f_0 X $ and $ \lim_{X\rightarrow 0} f(X) = f_1 X^{3/2}$
 we find that the Newtonian limit fixes the Newton's constant
$G_N = \left(\frac{2}{f_0} + \frac{2}{2-K}\right)G$ and gives the PPN parameter $\gamma = 1$. In the
 MOND limit, we find that Milgrom's constant $a_0$ is given by $a_0 = \frac{2}{3 f_1} \frac{1}{\frac{1}{f_0} + \frac{1}{2-K}} e^{\phi_c} \frac{1}{\ell_0}$.

We find the field equations assuming isotropic coordinates. We assume that the vector field has no spatial component. 
The Einstein metric is given by $\tilde{g}_{00} = -e^{2\tilde{\Phi}}$ and $\tilde{g}_{ij} = e^{2\tilde{\Psi}}\gamma_{ij}$.
The unit-timelike constraint $A_a A^a=-1$
then fully determines the  vector field as $A_a = (e^{\tilde{\Phi}},\vec{0})$ and $A^a = (-e^{-\tilde{\Phi}},\vec{0})$. 
We further find that $\Phi = \tilde{\Phi}+\phi$ and $\Psi = \tilde{\Psi}-\phi$.

With this in hand, the vector field equations are identically satisfied.
The vacuum scalar field equation for our metric ansatz is unchanged from the AQUAL case, while the Einstein equations become
\begin{eqnarray}
&&
  2\tilde{\Psi}''  + \tilde{\Psi}' \left(\tilde{\Psi}' + \frac{4}{r}  \right) =     -\frac{1}{2\ell_0^2} f  e^{2\tilde{\Psi}} 
-  K\left[    \tilde{\Phi}''
+    \tilde{\Phi}' \left( \frac{1}{2}\tilde{\Phi}' +\tilde{\Psi}' +  \frac{2}{r} \right)   
\right]
\qquad
\label{G_tt_T}
\\
&&
 \frac{2}{r}\left(\tilde{\Phi}'  +\tilde{\Psi}'\right)  +  \tilde{\Psi}'\left(2\tilde{\Phi}'  + \tilde{\Psi}'\right)
 = 
\frac{1}{\ell_0^2}   e^{2\tilde{\Psi}}\left( X f_X  - \frac{1}{2} f \right)
 - \frac{1}{2}  K \tilde{\Phi}'{}^2 
\label{G_rr_T}
\\
&&
 \tilde{\Phi}''+ \tilde{\Psi}''  +  \frac{1}{r}\left(\tilde{\Phi}'  +\tilde{\Psi}'\right) + (\tilde{\Phi}')^2  
 = - \frac{1}{2\ell_0^2} f  e^{2\tilde{\Psi}}  + \frac{1}{2} K  \tilde{\Phi}'{}^2
\label{G_AB_T}
\end{eqnarray}
In the MG limit we have that $f=f_1 X^{3/2}$, hence the scalar field equation can be integrated as in the AQUAL case. Once again,
in the $r\rightarrow \infty$ limit, the scalar-dependent terms in the Einstein equations vanish more rapidly than the metric terms and the
corresponding field equations become those of Einstein-{\AE}ther theory. The solutions in that case have been found by Eling and Jacobson~\cite{ElingJacobson2006}.
We do not need the exact solution here but only the fact that as  Eling and Jacobson show, the solution is asymptotically flat. We may then expand 
 $\tilde{\Phi} = \frac{\Phi_1}{r} + \ldots$ and $\tilde{\Psi} = -\frac{\Phi_1}{r} + \ldots$, while
$\phi = c \ln \frac{r}{r_0}   +  \frac{c\Phi_1}{2r} + \ldots$ where $c>0$ is given as before. This gives the physical metric as
\begin{eqnarray}
ds^2 &=&- \left(\frac{r}{r_0}\right)^{2c}  \left(1 + \frac{(2+c)\Phi_1}{r} + \ldots \right)   dt^2 
\nonumber 
\\
&& \ \ \ \
+  \left(\frac{r}{r_0}\right)^{-2c} \left(1 -\frac{(2+c)\Phi_1}{r} + \ldots\right)  dL^2 
\end{eqnarray}
which is the metric (\ref{met_T}) in the limit $r \rightarrow \infty$.

It is instructive once again to find the physical metric in  Schwarzschild  coordinates $\bar{r}$. Assuming for the  moment that $c<1$ (we shall motivate this further below), 
the coordinate transformation in the $r\rightarrow \infty$ limit implies
that $\frac{\bar{r}}{r_0} =  \left(\frac{r}{r_0}\right)^{1-c}    \left[1 - \frac{(2+c)\Phi_1}{r} + \ldots\right]$ which can be inverted to give
$\frac{r}{r_0} = \left(\frac{\bar{r}}{r_0}\right)^{\frac{1}{1-c}}\left[1 + \frac{(2+c)\Phi_1}{2(1-c)r_0} \left(\frac{r_0}{\bar{r}}\right)^{\frac{1}{1-c}} + \ldots\right]$,
therefore $\bar{r}\rightarrow \infty$ as $r\rightarrow \infty$. 
Thus we find the metric in  Schwarzschild  coordinates as 
\begin{equation}
ds^2 =
 -   \left(\frac{\bar{r}}{r_0}\right)^{\frac{2c}{1-c}} dt^2
+ \frac{   d\bar{r}^2 }{(1-c)^2}
+ \bar{r}^2  d\Omega^2 
\end{equation}
As in the case of AQUAL, the spatial metric is of the same form as the Barriola-Vilenkin global monopole solution~\cite{BarriolaVilenkin1989},
the factor $\frac{1}{1-c}$ appearing in the $d\bar{r}$ part of the metric implying now a deficit solid angle given by $1-c$. 

 It is interesting that as $c\rightarrow 1$ the transformation becomes singular. Does it even make sense if $c>1$? We can answer this by first looking at the curvature invariants.
The  curvature tensors go as
$R^{ab}_{\phantom{ab}cd} =  P^{ab}_{\phantom{ab}cd} r^{2(c-1)}$,
$R^{a}_{\phantom{a}b} =  P^{a}_{\phantom{a}b}  r^{2(c-1)} $
$R = 2(1-c) r^{2(c-1)}$, where the $P$-tensors are constant.
So forming invariants by contracting various types of curvature tensors together, or their derivatives, we would get that a typical invariant would be
$I_{n,d} = \frac{J}{r^{2n(1-c)+d}} $, where $J$ is a real number, $n$ is the number of curvature tensors involved
and $d$ is the number of derivatives acting on the curvature tensors. Thus all curvature invariants vanish as
$r\rightarrow \infty$ provided $c<1$ and if $c>1$ they diverge. This indicates that
$c\ge 1$ is not physical. Another way to see this is as follows. The proper distance is $D = \frac{r_0}{c-1} \left(\frac{r}{r_0}\right)^{1-c}$, hence,
for $c>1$ we find that $D\rightarrow 0$ as $r\rightarrow \infty$. Therefore, if the solution was valid for $c>1$  there would be a naked singularity at zero
proper distance, surrounded by MONDian spacetime up-to some finite distance $r_0$.
We conclude that the case $c\ge1$ is not physical in TeVeS theory.

\subsection{ Generalized Einstein {\AE}ther gravity}
The Generalized Einstein Aether theory involves a unit-timelike vector field much like TeVeS, however, it has no scalar field.
The theory is formulated with a single metric, and the MOND behaviour is recovered  by using a 
 noncanonical kinetic term \cite{ZlosnikFerreiraStarkman2007} for the vector field. Its action is as follows:
\begin{eqnarray}
 S [g^{ab},A^{a},\lambda] &=&  \frac{1}{16\pi G} \int d^4x \sqrt{-g} \bigg[R   - \frac{1}{\ell^2}  f(X) 
\nonumber 
\\
&& 
\ \ \ \ + \lambda \left(A^a A_a + 1 \right) \bigg] 
       + S_{\mbox{matter}}[g]
\label{geaact}
\end{eqnarray}
where 
\begin{equation}
  X = \ell^2 K^{abcd} \nabla_a A_b \nabla_c A_d
\end{equation}
and $ K^{abcd}$ is given by (\ref{K_ten}) with $\tilde{g}\rightarrow g$.

The theory depends on a free function  $f(X)$ which must have appropriate limits to recover Newtonian or MOND gravity in the non-relativistic case.
The non-relativistic limit of the theory gives back the MOND equation (\ref{eq:mond}) 
with $\mu =    \frac{2 - K f_X}{2 - K}$, where $K=c_1-c_4$. The Newtonian limit is recovered  when $\lim_{X\rightarrow\infty}f(X) =  X $  
in which case
the observed Newton's constant is fixed as $G_N = \frac{2}{2-K}G$ and the  PPN parameter $\gamma = 1$.
To get MOND behaviour we define a new function $h(X) =  f(X) - \frac{2X}{K}$, so that  $\mu =   -\frac{Kh_X}{2 - K}$. MOND is recovered if 
 $ \lim_{X\rightarrow 0} h(X) = f_1 (-X)^{3/2}$ 
in which case we get that Milgrom's constant $a_0$ is given by $a_0 = \frac{2(2-K)}{3 f_1 K^{3/2}}   \frac{1}{\ell_0}$.

We now proceed to find the relativistic MOND solution.  We assume that there is no spatial tilt to the vector field $A^{a}$. Consequently, 
the vector field is entirely fixed in terms of the metric by the fixed-norm constraint and the vector field equations are identically satisfied.
After calculation it may be shown that the vacuum Einstein equations reduce to the following: 
\begin{equation}
\label{ein1}
(h_X \Phi')' + h_X \Phi' \left[ \Phi' + \Psi' + \frac{2}{r} \right] + \frac{e^{2\Psi}}{K \ell_0^2 } \left( X h_X - h \right) = 0
\end{equation}
and
\begin{equation}
\label{ein2}
 2 \left( \Phi'' + \Psi''\right) + \frac{4}{r} ( \Phi' + \Psi') + (\Phi' + \Psi')^2 =  \frac{e^{2\Psi}}{K\ell_0^2 } \left( X h_X - \frac{3}{2} h \right)
\end{equation}
Substituting $h= f_1 (-X)^{3/2}$ for pure MOND, where $X = -  \ell_0\sqrt{K} e^{-\Psi} \Phi'$, the equations become:
\begin{equation}
\Phi'' +  \Phi' \left[ \frac{1}{3}\Phi'  + \frac{1}{r} \right]  = 0
\end{equation}
and
\begin{equation}
 Y''   + \frac{2}{r} Y' + \frac{1}{2} {Y'}^2=0
\end{equation}
where  $Y = \Phi+\Psi$. The solution is
\begin{eqnarray}
\Phi &=& \Phi_0 +  3 \ln \left( 1 + \frac{c}{3} \ln\frac{r}{r_0}\right)
\\
Y &=& Y_0 + 2\ln\left(1-\frac{1}{2}\frac{r_{0}}{r}\right)
\end{eqnarray}
where we have the constants of integration $c$,$\Phi_{0}$, $r_0$, and $Y_0$. As can be seen in section (\ref{circular}), the
number $c$ has a clear physical interpretation. The numbers $\Phi_{0}$ and $Y_0$ may without loss of generality be set to zero by rescale the temporal and radial coordinates.
We this arrive at the metric
\begin{eqnarray}
ds^2_{\AE} &=& -\left[1+\frac{c}{3}\ln\left(\frac{r}{r_{0}}\right)\right]^{6}dt^{2} 
+ \frac{\left(1-\frac{r_0}{2r}\right)^2}{\left[1+\frac{c}{3}\ln\left(\frac{r}{r_{0}}\right)\right]^6 } dL^2
\end{eqnarray}
For large distances $r\gg r_0$ we recover the metric (\ref{met_GEA}).

The curvature tensors go as $R^{ab}_{\phantom{ab}cd} = \frac{1}{r^2} P^{ab}_{\phantom{ab}cd}(\ln \frac{r}{r_0})$
$R^{a}_{\phantom{a}b} = \frac{1}{r^2} P^{a}_{\phantom{a}b}(\ln \frac{r}{r_0})$
$R = \frac{1}{r^2} P(\ln \frac{r}{r_0})$, where the $P$-tensors are polynomial functions of $\ln \frac{r}{r_0}$.
So forming invariants by contracting various types of curvature tensors together, or their derivatives we would get that a typical invariant would be
$I_{n,d} = \frac{1}{r^{2n+d}} J(\ln \frac{r}{r_0})$, where $J$ is a polynomial function of $\ln \frac{r}{r_0}$, $n$ is the number of curvature tensors involved
and $d$ is the number of derivatives acting on the curvature tensors. Thus all curvature invariants vanish as
$r\rightarrow \infty$.

Finally we note some novel behaviour at the point $X\rightarrow 0$. Here the Einstein equation (\ref{ein1}) is identically zero leaving only equation equation (\ref{ein2})
which becomes again a differential equation for the variable $Y=\Phi+\Psi$. It may also be shown that the action (\ref{geaact}) in this limit is only a function of $Y$ and is thus
invariant under $\Phi \rightarrow \Phi + \beta(r)$ and $\Psi \rightarrow \Psi -\beta(r)$. 
 This disformal symmetry can be more explicitely written as
$g_{ab}\rightarrow e^{-2\beta}g_{ab}-\sinh(2\beta)A_{a}A_{b}$ and $A^{a}\rightarrow e^{-\beta}A^{a}$ where $\beta(r)$ is some function of $r$.
Interestingly the metric transformation is of the same form as the transformation between Einstein and physical metrics in the TeVeS model \cite{Bekenstein2004a}.

\section{Timelike Geodesics}

\subsection{General motion of test particles on planar orbits}
We now seek to look at effects which do not rely on the nonrelativistic conditions being safistied.

Consider a test particle's four velocity $U^{a}$. In terms of the proper time $\tau$ along the trajectory we have $U^{a}=dx^{a}/d\tau\equiv \dot{x}^{a}$. Furthermore, without loss of generality we may consider trajectories lying in $\theta=\pi/2$.
Hence, the unit-timelike condition on the particle's four velocity may be expressed as follows:

\begin{equation}
-1 = -e^{2\Phi}\dot{t}^{2}+e^{2\Psi}(\dot{r}^{2}+r^{2}\dot{\varphi}^{2})
\end{equation}
Furthermore we have some constants of the motion. The first is the projection of the four velocity along the timelike Killing vector $(\partial_{t})^{a}$:
\begin{equation}
 E \equiv  -g_{ab}(\partial_{t})^{a}U^{b} = e^{2\Phi}\dot{t}
\end{equation}
The second is the projection of the four velocity along the angular Killing vector $(\partial_{\varphi})^{a}$:
\begin{equation}
L \equiv  g_{ab}(\partial_{\varphi})^{a}U^{b} = e^{2\Psi}r^{2}\dot{\varphi}
\end{equation}

These expression may then be put in the unit-timelike condition $U^{a}U^{a}g_{ab}=-1$ obeyed 
by the test particle, which after rearranging becomes
\begin{equation}
\frac{E^{2}}{2}=\frac{1}{2}e^{2\Phi+2\Psi}\dot{r}^{2}+\frac{e^{2\Phi}}{2}+\frac{L^{2}}{2r^{2}}e^{-2\Psi+2\Phi}
\end{equation}

We now define the new variable $\xi \equiv \int e^{\Psi+\Phi}dr$, in terms of which the energy equation becomes:

\begin{eqnarray}
{\cal E} &\equiv & \frac{E^{2}}{2} = T + V \\
T  &\equiv & \frac{1}{2}\dot{\xi}^{2} \\
V &\equiv & \frac{e^{2\Phi}}{2}+\frac{L^{2}}{2r^{2}(\xi)}e^{-2\Psi+2\Phi}
\end{eqnarray}

Hence, the problem of motion of test particles is analogous to a one dimensional classical mechanics problem of a particle with energy per unit mass $E^{2}/2$, kinetic energy per unit mass $T$, and potential energy per unit mass $V(\xi)$.

\subsection{Circular Orbits}
\label{circular}

In keeping with our classical mechanics analogy, circular orbits will exist given appropriate initial data and the existence of a value of $\xi$ which satisfies $dV/d\xi=0$. Assuming the existence of such a point, we may relate the angular momentum to the geometry at that point, leading to the following relation:
\begin{equation}
L^{2} = e^{2\Psi}\frac{\frac{d\Phi}{d\xi}r^{3}}{\frac{dr}{d\xi}+r\left(\frac{d\Psi}{d\xi}-\frac{d\Phi}{d\xi}\right)}
\end{equation}
We can express this in terms of a more familiar quantity, the norm of the spatial part of the particle's four velocity, defined as  $v^{2} \equiv h_{ab}U^{a}U^{b}$ where $h_{ab}$ is the spatial metric.  It may be checked that $L^{2}=v^{2}r^{2}e^{2\Psi}$ and so
\begin{equation}
v^{2}=\frac{\frac{d\Phi}{d\xi}r}{\frac{dr}{d\xi}+r\left(\frac{d\Psi}{d\xi}-\frac{d\Phi}{d\xi}\right)}
\end{equation}
Clearly then if $e^{2\Phi}$ and $e^{2\Psi}$ are monomial functions of $r$ then $v$ will asymptote to a constant. This a fully relativistic
realization of the so-called Tully-Fisher relation that MOND reproduces. In AQUAL we have $v^{2}=c=\sqrt{GMa_{0}}$. In TeVeS we have
$v^{2}=c/(1-2c)$. As we shall see, this result is to be expected: for $c>(1/2)$ there exists no minimum of the potential $V(\xi)$ and hence
there cannot exist any circular orbits. Finally, we see that although the GEA model reproduces the TeVeS result in the nonrelativistic limit,
this eventually becomes a poorer approximation and $v^{2}$ is no longer approximately constant across increasing radii of orbits. This transition  will happen when $\ln(\xi/r_{0})$ becomes of order unity, representing a significant deviation of $\Phi'$ and $\Psi'$ from power law behaviour. By inspection this will happen at a radius we will call $r_{a}$ which is found to be:
\begin{equation}
r_{a}= r_{0}e^{\frac{3}{c}}
\end{equation}

This is a novel feature. The AQUAL and TeVeS metrics maintain characteristically `MOND' behaviour to infinitely large distances from the source, as is evidenced by $v^{2}$ 
remaining constant as one varies orbital radius. This is not the case however in the GEA model, which approximates TeVeS only in the regime $r_{0} \ll r \ll r_{a}$. The 
extent of this regime depends on the value of $c=\sqrt{GMa_{0}}$ and thus the mass of the sourcing object. By way of example, the value of $c$ exterior to a galaxy 
such as the Milky Way is of order $10^{-6}$, meaning that here $r_{a}$ is a vastly greater value than the radius at onset of the modified gravity regime.

\subsection{The Precession of almost-circular Orbits}

We now seek to generalize to examine the properties of orbits that are not circular. 
Differentiating the energy equation we have that:
\begin{equation}
\ddot{\xi}=-\frac{d V}{d\xi}
\end{equation}
In order to deduce how the orbital radius varies with angle, we would next like to convert $\ddot{\xi}$ into $d^{2}\xi/d\varphi^{2}$.
From the conservation of angular momentum we have $\frac{d}{d\tau}= \frac{Le^{-2\Psi}}{r^{2}}\frac{d}{d\varphi}$, therefore,
\begin{equation}
\ddot{\xi} = \frac{L^{2}e^{-2\Psi}}{r^{2}}\frac{d}{d\varphi}\left( \frac{e^{-2\Psi}}{r^{2}}\frac{d\xi}{d\varphi}\right)
\label{ddot_xi}
\end{equation}
For simplicity of calculation we further define a new variable $u=-\int (e^{-2\Psi}/r^{2})d\xi$ and in terms of $u$ rather than $\xi$, the orbital equation becomes
\begin{equation}
\label{orb1}
\frac{d^{2}u}{d\varphi^{2}} = -\frac{1}{L^{2}}\frac{dV}{du}
\end{equation}

A strategy for showing whether orbits are closed is to look at almost circular orbits. i.e. trajectories with $u=u_{p}+u_{0}$ where $u_{0}$ is
the function evaluated at the radius of a circular orbit and $u_{p}$ is small in some suitable sense. In this situation we may look to Taylor expand the right hand side
of equation (\ref{orb1}) in powers of $u_{p}$. The assumed smallness of $u_{p}$ means that we can, to a good approximation, neglect terms in $u_{p}^{2}$ and higher orders.
The result is an equation where the second derivatives of $u_{p}$ with respect to $\varphi$ are proportional to $u_{p}$. The solutions to this equation
thus be $u_{p}=u_{p0}\cos(\Omega\varphi+b)$ where $u_{p0}$ and $b$ are constants of integration. If $\Omega$ is an integer then almost circular orbits are closed.

\subsubsection{General Relativity}

For comparative purposes, we first demonstrate this calculation for the Schwarzschild solution of General Relativity. In isotropic co-ordinates the metric takes the following approximate form:
\begin{equation}
\label{grmet}
ds^2_{GR}= -\left[1-2\Phi_N + 2 \Phi_N^2+ \ldots\right]dt^{2} + \left[1+2\Phi_N+ \ldots\right]dL^2
\end{equation}
where $\Phi_N = \frac{GM}{r}$. One may proceed with the method described above to obtain a solution of the form $u_p = u_{p0} \cos(\Omega_{GR}\varphi + b)$, where
there is a one-to-one correspondence between the field $u_p(r)$ and $r$. The quantity $\Omega_{GR}$ is found to be equal to $\sqrt{1 - \frac{6GM}{r_c}} \approx 1 - \frac{3GM}{r_c}$,
where $r_c$ is the radius of the circular orbit that is perturbed around. Therefore, $u_p$ has an angular period of approximately
$2\pi (1 + \frac{3GM}{r_c})$ radians. 
Significantly, the deviation from closedness decreases as the corresponding circular orbital radius increases.

\subsubsection{AQUAL}
Recall that the asymptotic metric in AQUAL is given by (\ref{met_A}). 
Therefore, $\xi$ is not identical to $r$ but is rather $\xi=\int e^{2\Phi} dr $. 
However, there is a compensating simplification in the relation of the variable $u$ to the original coordinates:
\begin{equation}
u \equiv -\int \frac{e^{-2\Psi}}{r^{2}}d\xi = -\int \frac{e^{2\Psi-2\Phi}}{r^{2}}dr= \frac{1}{r}
\end{equation}
It is then simple to write $V$ as a function of $u$:
\begin{equation}
V= \frac{1}{2}\left(\frac{1}{ur_{0}}\right)^{2c}+\frac{1}{2}L^{2}u^{2}
\end{equation}
The solution for $u_{0}$ is found to be:
\begin{equation}
u_{0} =\frac{1}{r_{0}} \left(\frac{L^{2}}{c r_{0}^{2}}\right)^{-\frac{1}{2(c+1)}}
\end{equation}
After some straightforward algebra one finds that:
\begin{equation}
\frac{d^{2}u_{p}}{d\varphi^{2}} = -2(1+c)u_{p}
\end{equation}

We see then that the angular frequency of $u_{p}$ in AQUAL is given by  $\Omega_{A}^{2}=2(1+c)$.  Therefore, near circular orbits
in AQUAL are not closed unless the number $c$ is an integer. Note also that the angular period is significantly different from 
$2\pi$, even if $c\ll 1$ and in the general case of non-integer $c$, the deviation from closedness is independent of orbital radius.

\subsubsection{TeVeS}
We recall that at far distances from the source, the metric in TeVeS is given by (\ref{met_T}).  As in General Relativity, $\xi$ may be simply identified with $r$. 
We first find the variable $u$. It is found to be defined via the equation:
\begin{equation}
u= \frac{1}{(1-2c)r_{0}}\left(\frac{\xi}{r_{0}}\right)^{2c-1}
\end{equation}
Meanwhile the potential $V$ may be expressed in terms of $u$ simply as:
\begin{equation}
V = \frac{1}{2}\left[(1-2c)r_{0}u\right]^{\frac{2c}{2c-1}}+\frac{L^{2}}{2}(1-2c)^{2}u^2
\end{equation}
As before we find $u_{0}$, the value that $u$ takes at the circular orbit. This is found to be as follows:
\begin{equation}
u_{0}=\frac{1}{(1-2c)}\frac{1}{r_{0}}\left(\frac{(1-2c)L^{2}}{cr_{0}^{2}}\right)^{\frac{2c-1}{2(1-c)}}
\end{equation}
After a straightforward but lengthy calculation it may be shown that the perturbation $u_{p}$ obeys the following equation:
\begin{equation}
\frac{d^{2}u_{p}}{d\varphi^{2}} = -2(1-2c)(1-c) u_{p}
\end{equation}

Therefore the frequency is given by $\Omega_{T}^{2}=2(1-2c)(1-c)$, which will typically be markedly different from integer values.
As in AQUAL, the angular frequency differs significantly from $2\pi$ even if $c\ll1$.
It is noteworthy that for the range $1/2<c<1$, the perturbation $u_{p}$ will grow exponentially. Indeed, by inspection the potential $V(\xi)$ can no
longer have a minima in this case, and will inevitably push the test particle to smaller values of $\xi$. Recalling that $c=\sqrt{GM a_{0}}$, we see that in TeVeS
if there exists an object whose mass  is greater than $1/(4a_{0}G)$ and has a region exterior to it in the deep MOND regime, then this region should be expected to be devoid of matter.
Similar conclusions will hold in the GEA model as long as the condition $r\ll r_{a}$ is satisfied.

\subsection{Radial Trajectories}

We now ask whether it is possible for a test particle to be projected to spatial infinity. The method to do this with the least energy would clearly be to do so with
vanishing angular momentum $L$ as this will only detract from how much of the energy ${\cal E}$ can go into the radial motion of the particle.

\subsubsection{General Relativity}
Recalling the metric (\ref{grmet}) we may immediately write down the zero angular momentum energy equation:
\begin{equation}
{\cal E} = \frac{1}{2}\dot{\xi}^{2}+\frac{1}{2}\left(1-\frac{2GM}{\xi}\right)
\end{equation}

Consider a test particle propelled radially outwards from position $\xi_{0}$ and with velocity $(\dot{\xi})_{0}$. For the trajectory to reach $\xi=\infty$ 
having slowed down to $\dot{\xi}=0$, it must have energy ${\cal E}=1/2$, and of course is the least amount of energy necessary to reach spatial infinity. Additionally this implies that we have the relation $(\dot{\xi})_{0}=\sqrt{2GM/\xi_{0}}$

\subsubsection{AQUAL}
In the AQUAL case, although $\xi$ is no longer identical to $r$, it may readily be checked that $\xi=\infty$ indeed corresponds to spatial infinity for all positive values of $c$. The zero angular momentum energy equation takes the form:
\begin{equation}
{\cal E} = \frac{1}{2}\dot{\xi}^{2}+\frac{1}{2}\left(\frac{\xi}{r_{0}}\right)^{2c}
\end{equation}
We immediately see that there exists no finite value of ${\cal E}$ for which the particle may begin at a finite value $\xi_{0}$ and reach $\xi=\infty$, or equivalently, there is no finite initial velocity with which the particle could be propelled to each spatial infinity.

\subsubsection{TeVeS}

Recall that $\xi$ is identically equal to $r$. The proper distance $D$ between $\xi=\xi_{0}$ and $\xi=\infty$ is then:
\begin{equation}
D= \int^{\infty}_{\xi_{0}}\left(\frac{\xi}{r_{0}}\right)^{-c}d\xi
\end{equation}
Since $c< 1$ in TeVeS, these co-ordinates span an infinite proper distance. The zero angular momentum energy equation is then:
\begin{equation}
{\cal E} = \frac{1}{2}\dot{\xi}^{2}+\frac{1}{2}\left(\frac{\xi}{r_{0}}\right)^{2c}
\end{equation}
Precisely as in AQUAL, no finite ${\cal E}$ permits the test particle to reach spatial infinity. 

\subsubsection{GEA}
As in the case of TeVeS, $\xi$ may be identified with $r$. The proper distance between D between $\xi=\xi_{0}$ and $\xi=\infty$ is of the form:
\begin{equation}
D= \int^{\infty}_{\xi_{0}}\left(1+\frac{c}{3}\ln\left(\frac{\xi}{r_{0}}\right)\right)^{-3}d\xi
\end{equation}
This integral diverges for all positive values of $c$ and so isotropic coordinates are again appropriate.
\begin{equation}
{\cal E} = \frac{1}{2}\dot{\xi}^{2}+\frac{1}{2}\left(1+\frac{c}{3}\ln\left(\frac{\xi}{r_{0}}\right)\right)^{6}
\end{equation}
Once again, no finite ${\cal E}$ permits the test particle to reach spatial infinity. Therefore, for all of the spacetimes considered here, there can exist no unbound orbits.
Put more informally, in AQUAL, TeVeS, and GEA the escape velocity for all massive particles is the speed of light.

\section{Null Geodesics}

We may assume that the trajectories of test photons follow null geodesics i.e. their wave vector $k^{a}=dx^{a}/d\lambda$ (where $\lambda$ is an affine parameter) satisfies $k^{a}k_{a}=0$. As before, we may make use of constants of the motion $E$ and $L$ which, as the notation suggests, correspond to constants of the motion associated with the projection of $k^{a}$ along the time like and $\partial_{\varphi}$ directed Killing vectors respectively. The null norm condition is then simply:
\begin{equation}
E^{2} = e^{2\Psi+2\Phi}\dot{r}^{2}+\frac{L^{2}}{r^{2}}e^{2\Phi-2\Psi}
\end{equation}
Therefore again we may make the analogy with a one dimensional classical mechanics problem of a test particle with a trajectory $\xi(\lambda)\equiv \int e^{\Psi+\Phi}dr$. Its energy equation is then  ${\cal E} = T+V$ with
\begin{eqnarray}
{\cal E} &\equiv & \frac{E^{2}}{2} \\
T &\equiv &  \frac{1}{2}\dot{\xi}^{2} \\
V & \equiv &\frac{1}{2}\frac{L^{2}}{r^2(\xi)}e^{2\Phi-2\Psi} 
\end{eqnarray}

\subsection{Gravitational Redshift}

We now consider the effect on a photon's frequency as it passes through the spacetime. We consider two stationary observers, one at coordinate $r_{1}$ and another at $r_{2}$ where $r_{2}>r_{1}$. Their four velocity $U^{a}(r)$ may be related to the spacetime's timelike Killing vector $\xi^{a}=(1,0,0,0)$ as follows:

\begin{equation}
\label{kill}
U^{a} = \frac{1}{\sqrt{-g_{ab}\xi^{a}\xi^{b}}}\xi^{a}
   =  \frac{1}{\sqrt{-g_{00}}}\xi^{a}
\end{equation}

Consider a photon trajectory with wavevector $k^{a}$  emitted at $r_{1}$ and absorbed at $r_{2}$. At $r_{1}$, the frequency of the photon
measured by the stationary observed is simply $\omega_{1}=-k^{a}U_{a}(r_{1})$. Similarly, the frequency of the photon measured 
later by the observer at $r_{2}$ is given by $\omega_{2}=-k^{a}U_{a}(r_{2})$. We may express the ratio of these frequencies in terms 
of the wavevector an the Killing vector using equation (\ref{kill}), yielding:

\begin{equation}
\frac{\omega_{2}}{\omega_{1}}= \frac{\sqrt{-g_{00}(r_{1})}}{\sqrt{-g_{00}(r_{2})}}\frac{k^{a}\xi_{a}(r_{2})}{k^{a}\xi_{a}(r_{1})}
\end{equation}

However, from the geodesic equation and Killing's equation, it follows that ${\cal L}_{k^{b}}(\xi_{a}k^{a})=0$ and so the ratio
$\frac{k^{a}\xi_{a}(r_{2})}{k^{a}\xi_{a}(r_{1})}$ is simply equal to unity. Thus the ratio of frequencies depends only on the varying norm of the Killing vector.

\subsubsection{TeVeS and AQUAL}

As the time-time component of the metric is identical in TeVeS and AQUAL, they will describe identical gravitational redshift. The result may simply 
be read off to be:
\begin{equation}
\left(\frac{\omega_{2}}{\omega_{1}}\right)_{A/T}=\left(\frac{r_{1}}{r_{2}}\right)^{c}.
\end{equation}

\subsubsection{GEA}
Again we may read off the result given the metric:
\begin{equation}
\left(\frac{\omega_{2}}{\omega_{1}}\right)_{GEA}=\left(\frac{1+\frac{c}{3}\ln\left(\frac{r_{1}}{r_{0}}\right)}{1+\frac{c}{3}\ln\left(\frac{r_{2}}{r_{0}}\right)}\right)^{3}
\end{equation}

Clearly in all three models, $\omega_{2}/\omega_{1}\rightarrow 0$ as $r_{2}\rightarrow \infty$.
Therefore, no photon can reach infinity with non-zero energy. This an  equivalent result to the case of massive particles.

\subsection{The Deflection Of Light}

Although the frequency of a photon diminishes to zero as it approaches infinite distance from the source, null geodesics can be extended to these distances 
for any positive value of ${\cal E}$. 
For each of the models, $V(\xi)\rightarrow 0$ as $\xi\rightarrow \infty$. We now consider a photon with finite and positive ${\cal E}$ which approaches the source from $\xi=\infty$, reaches a minimum distance from the source at $\xi_{min}$ and then returns to $\xi=\infty$. 
In a general spacetime the deflection angle is given by
\begin{equation}
\Delta = 2\int_{\xi_{min}}^{\infty}\frac{d\varphi}{d\xi} d\xi -\pi
\end{equation}

\subsubsection{AQUAL}

From the definition of $L$, as before, we may obtain an expression for $d\xi/d\varphi$, which turns out to be:
\begin{eqnarray}
\label{turn}
\frac{d\xi}{d\varphi} = \frac{e^{2\Phi}L\sqrt{2({\cal E}-V)}}{2V}
\end{eqnarray}
The deflection angle is then given by
\begin{equation}
\Delta = 2\int _{r_{min}}^{\infty}\frac{L}{r^{2}\sqrt{2({\cal E}-\frac{L^{2}}{2r^{2}})}}dr -\pi 
\end{equation}
The radius $r_{min}$ of closest approach is simply found from the condition that $\dot{\xi}=0$ at that point, and is found to be $r_{min}=\sqrt{L^{2}/(2{\cal E})}$
Evaluation of the above integral yields $\Delta=0$. This is to be expected. The asymptotic AQUAL metric is conformally flat, and it is well known that the lensing 
of photons is insensitive to conformal rescalings of the metric. It is this property that leads to substantial conflict with data as the deflection of light around galaxies
is known to strongly indicate the presence of mass discrepancies. The conditions under which a hypothetical \emph{purely metric} theory of gravity underlying MOND
would lead to this problem with the asymptotic metric have been explored in detail in \cite{Soussa:2003sc}.

\subsubsection{TeVeS}

We may proceed as before. The equation for $d\xi/d\varphi$ again takes the form (\ref{turn}). However, unlike the AQUAL case, $\xi$ is identical to $r$ and
the functional forms of $e^{2\Psi}$ differ.

The deflection angle is then
\begin{equation}
\Delta =2\int _{r_{min}}^{\infty}\frac{L}{r^{2}}\left(\frac{r}{r_{0}}\right)^{2c}\frac{1}{\sqrt{2\left({\cal E}-\frac{L^{2}}{2r^{2}}\left(\frac{r}{r_{0}}\right)^{4c}\right)}}dr -\pi 
 \end{equation}
Where we have found $r_{min}$ using the previously described approach, yielding: $r_{min}= r_{0}\left(2r_{0}^{2}{\cal E}/L^{2}\right)^{\frac{1}{2(2c-1)}}$.
For evaluation of the integral it is convenient to introduce the new variable $y=L/(r_{0}\sqrt{2{\cal E}})(r/r_{0})^{2c-1}$. In terms of this the integral becomes:
\begin{eqnarray}
\Delta &=& \frac{2}{2c-1}\int_{y_{min}}^{y_{max}}\frac{1}{\sqrt{1-y^{2}}}dy-\pi 
\nonumber
\\
         &=& \frac{2}{2c-1}\left[\sin^{-1}(y)\right]^{y_{max}}_{y_{min}}-\pi
 \end{eqnarray}

If $c<1/2$ then $y_{max}=0$, $y_{min}=1$, in which case we have that $\Delta=2c\pi/(1-2c)$. Therefore the deflection angle is nonzero and constant across different values of $r_{min}$. This can be compared to the velocity of circular orbits and precession rate of almost circular orbits which also are independent of the 
distance from the source characteristic to the problem. If $c=1/2$ then $V(\xi)$ becomes position
independent and so there is nothing to prevent the photon falling towards smaller radii than the modified gravity regime covers. If $1>c\geq 1/2$ then via our classical mechanics analogy $V(\xi)$ no longer represents a repulsive force and so
there is simply nothing to stop a photon impinging from $\xi=\infty$ from heading to smaller values of $\xi$ than are covered
by the asymptotic metric- i.e. the photon inevitably leaves the modified gravity regime.

\section{Conclusions}

In this article we have considered the properties of timelike and null geodesics for a set of metrics.
Though these metrics follow from specific models, we anticipate that they display features which are not merely confined to those models. It is 
conceivable that any of the metrics may follow from a conceptual framework that does not even involve a modified Poisson equation in the transition
from General Relativity to modified gravity regime.

By considering a fully relativistic treatment of possible asymptotic geometries for a theory of modified Newtonian dynamics,
we have found counterparts to the more familiar `flat rotation curve' property of the geometry-  namely that the deflection angle
for null geodesics and the precession rate of near circular orbits may additionally become independent of distance from the
source. Furthermore, we have in the case of TeVeS that the `scale invariance' inherent to the asymptotic geometry leads to a sensitivity to mass
that does not exist in General Relativity, specifically that exotic behaviour emerges  whenever $c=\sqrt{G_N Ma_{0}}\geq 1$, for any value $r\gg r_{0}$,
namely that the spacetime exchibits a physical singularity as $r\rightarrow \infty$ for $c>1$.
 We interprete this result as saying that physical spacetimes in TeVeS exists only if $0\le \sqrt{G_N M a_0} < 1$.

A common property of all the geometries considered here is that they all forbid unbound orbits for massive particles (the escape velocity is the speed of light)
and all of them will lead to a total diminishment of a test photon's frequency as it reaches spacelike infinity.

Additionally, we have seen in the form of the GEA model that a consistent relativistic model of MOND exists which agrees with nonrelativistic
phenomenology but which differs markedly in its behaviour from the other models as one approaches a radius $r_{a}$. For a source of mass $M$, 
the ratio of this radius to the radius at which the nonrelativistic modified gravity
becomes dominant is of order $e^{3/\sqrt{GMa_{0}}}$. 
In the most general terms then it seems difficult then to anticipate the geometry of an underlying theory of MOND as $r\rightarrow \infty$.

\section{Bibliography}
\bibliographystyle{unsrt}
\bibliography{references}

\appendix

\end{document}